\documentclass[square,numbers,referee,sn-basic]{sn-jnl}% referee option is meant for double line spacing

\usepackage{graphicx}%
\usepackage{multirow}%
\usepackage{amsmath,amssymb,amsfonts}%
\usepackage{amsthm}%
\usepackage{mathrsfs}%
\usepackage[title]{appendix}%
\usepackage{xcolor}%
\usepackage{textcomp}%
\usepackage{manyfoot}%
\usepackage{booktabs}%
\usepackage{listings}%
\usepackage{physics}%
\usepackage{tikz,tikz-3dplot} % A package for high-quality hand-made figures.
\usetikzlibrary{} % To use the TikZ environment
\usetikzlibrary{3d,calc,arrows.meta,shapes.geometric}
\usepackage{tkz-euclide} % To draw 2D figures
\usetikzlibrary{
	decorations.pathreplacing,
	decorations.pathmorphing,
	calligraphy,
	patterns.meta,
	shapes
}
\usepackage[linesnumbered,ruled,vlined]{algorithm2e}
\definecolor{mygreen}{RGB}{0,128,0}
\definecolor{myblue}{RGB}{0,100,255}
\definecolor{myorange}{RGB}{255,140,0}
\definecolor{mygreen2}{RGB}{46,139,87}
\definecolor{rred}{RGB}{125,0,0}
\definecolor{ggreen}{RGB}{0,125,0}
\definecolor{bblue}{RGB}{0,0,125}
\definecolor{purple_uu}{RGB}{212,0,212}
\definecolor{blue_uu}{RGB}{16,0,213}
\definecolor{mybrown}{RGB}{181, 101, 29}

\begin{document}

\title[Article Title]{Deep reinforcement learning for the management of the wall regeneration cycle in wall-bounded turbulent flows}


\author*[1]{\fnm{Giorgio Maria} \sur{Cavallazzi}}\email{giorgio.cavallazzi@city.ac.uk}

\author[2,3]{\fnm{Luca} \sur{Guastoni}}\email{luca.guastoni@tum.de}
\equalcont{These authors contributed equally to this work.}
\author[3]{\fnm{Ricardo} \sur{Vinuesa}}\email{rvinuesa@mech.kth.se}
\equalcont{These authors contributed equally to this work.}
\author[1]{\fnm{Alfredo} \sur{Pinelli}}\email{alfredo.pinelli.1@city.ac.uk}
\equalcont{These authors contributed equally to this work.}

\affil*[1]{\orgdiv{Department of Engineering}, \orgname{City St George's, University of London}, \orgaddress{\street{Northampton Square}, \city{London}, \postcode{EC1V 0HB}, \state{} \country{UK}}}

\affil[2]{\orgdiv{School of Computation, Information and Technology}, \orgname{TU Munich}, \orgaddress{\street{Boltzmannstr. 3}, \city{Garching}, \postcode{D-85748}, \state{}  \country{Germany}}}

\affil[3]{\orgdiv{FLOW, Engineering Mechanics}, \orgname{KTH Royal Institute of Technology}, \orgaddress{\city{Stockholm}, \postcode{SE-100 44}, \country{Sweden}}}


\abstract{
The \textit{wall cycle} in wall-bounded turbulent flows is a complex turbulence regeneration mechanism that remains not fully understood. This study explores the potential of deep reinforcement learning (DRL) for managing the wall regeneration cycle to achieve desired flow dynamics. 
To create a robust framework for DRL-based flow control, we have integrated the \textit{StableBaselines3} DRL libraries with the open-source direct numerical simulation (DNS) solver \textit{CaNS}.
The DRL agent interacts with the DNS environment, learning policies that modify wall boundary conditions to optimise objectives such as the reduction of the skin-friction coefficient or the enhancement of certain coherent structures' features. 
The implementation makes use of the message-passing-interface (MPI) wrappers for efficient communication between the Python-based DRL agent and the DNS solver, ensuring scalability on high-performance computing architectures.

Initial experiments demonstrate the capability of DRL to achieve drag reduction rates comparable with those achieved via traditional methods, although limited to short time intervals. We also propose a strategy to enhance the coherence of velocity streaks, assuming that maintaining straight streaks can inhibit instability and further reduce skin friction.

Our results highlight the promise of DRL in flow-control applications and underscore the need for more advanced control laws and objective functions. Future work will focus on optimizing actuation intervals and exploring new computational architectures to extend the applicability and the efficiency of DRL in turbulent flow management.}


\maketitle

\section{Introduction}\label{sec1}

The wall cycle \citep{jimenezAutonomousCycleNearwall1999} in wall-bounded turbulent flows is a turbulence regenerating mechanism whose dynamics are still not fully understood. Understanding and controlling the wall-cycle is a technological challenge primarily pursued to either reduce the skin-friction coefficient 
or to achieve a desired impact on flow dynamics (e.g. enhancement of heat and mass transfer).

Active flow control techniques are among the most promising methodologies to manipulate the near wall flow for technological benefits because of 
their ability to work with closed-loop feedback and handle off-design conditions. However, their complexity currently limits real-world application. One such technique is based on the use of Streamwise Travelling Waves (STW) of spanwise velocity \citep{quadrioStreamwisetravellingWavesSpanwise2009}, which involves applying a sinusoidal wave with a time-dependent pulsation as a boundary condition for the spanwise velocity $w$ at the wall. The use of this method can lead to skin friction coefficient $C_f$ reduction by up to $45\%$ \citep{quadrioStreamwisetravellingWavesSpanwise2009}.
However, the definition of a particular STW leading to the highest drag reduction requires a computationally expensive 
analysis of the parametric space at hand, defined by the frequency and spatial wavenumber of the travelling wave

Deep Reinforcement Learning
\citep{lecunDeepLearning2015}\citep{liDeepReinforcementLearning2018} offers a potential breakthrough
in fluid mechanics applications, although its use in this area 
% by overcoming the limitations of standard parametric studies, enhancing existing flow control techniques, and enabling dynamic variations of the parameters. Despite these promising advances, the application of DRL to fluid mechanics 
is still in its early stages compared to traditional supervised learning approaches \citep{garnierReviewDeepReinforcement2021}\citep{rabaultArtificialNeuralNetworks2019}.
%[29,30]. 
Two primary categories of research contributions have emerged in this field: the first one focuses on using DRL for turbulence modelling \citep{novatiAutomatingTurbulenceModelling2021}, while the second applies DRL to active flow control. Our present work falls into the latter category, aiming to explore how DRL can be used to control complex flow systems dynamically and efficiently \citep{guastoniDeepReinforcementLearning2023}.

A few literature contributions looked at the DRL control of the movement of an object within a fluid flow. For instance, some studies represent a fish swimming in turbulent flow or within a school of fish, where the reward function is designed to maximise the swimming efficiency 
\citep{biferale2019zermelo,verma2018efficient}.
%[31,32]. 
Another group of studies leverages DRL to control the dynamics of the fluid itself. One prominent example of this is found in \citep{rabaultArtificialNeuralNetworks2019}, where the flow around a cylinder is controlled using jets perpendicular to the main flow direction. This setup has since been adopted as a benchmark in numerous studies \citep{rabault2019accelerating, paris2021robust, tang2020robust, fan2020reinforcement, ren2021applying, xu2020active},
%[33-38], 
highlighting both the growing interest in DRL for active flow control and the importance of developing benchmark cases to serve as a foundation for future research.

In this study, we use DRL as an active flow control technique for wall-bounded turbulent flows, a significant increase in complexity for two key reasons. Firstly, the two-dimensional (2D) cylinder case presents a well-defined vortex shedding pattern at a specific frequency, which dominates the flow. Suppressing this frequency offers a direct venue for controlling the wake. However, at higher Reynolds numbers, a different strategy is employed, where the agent energises the boundary layer on the cylinder’s surface to induce drag crisis \citep{varela2022deep}. In contrast, wall-bounded flows, such as those examined here, are inherently multi-scale phenomena, with no clear dominant frequencies or mechanisms to target for drag reduction.  Indeed, turbulent channel flows represent an inherently three-dimensional (3D) situation populated by turbulent structures, featuring a nonlinear, chaotic, and challenging benchmark for testing DRL algorithms and control strategies.

Recent efforts have begun to investigate similar flow cases, such as in \citep{sonodaReinforcementLearningControl2023}, where DRL was applied to a standard channel flow, and in \citep{zeng2022data}, where a Couette flow was controlled using two streamwise parallel slots. In the latter case, the DRL agent was trained on a reduced-order model before being applied to the full problem.
Another example, closely related to the present contribution, the work of Guastoni et al. \citep{guastoniDeepReinforcementLearning2023, walchli2024drag}, where the traditional opposition control method (i.e. blowing and suction from the wall) was improved by using a DRL algorithm, specifically deep a deterministic policy gradient (DDPG). The findings demonstrated that DRL could achieve up to 30\% drag reduction, surpassing the performance of the classical opposition control technique \citep{choiActiveTurbulenceControl1994}.

Finally, it is worth mentioning that DRL has been applied to a variety of other control tasks, spanning simple one-dimensional falling-fluid instabilities \citep{belus2019exploiting, vignonEffectiveControlTwodimensional2023}, convection problems \citep{beintema2020controlling} chaotic turbulent combustion systems \citep{bucci2019control}, and a range of engineering applications \citep{henry2022deep, zheng2021active, vinuesa2022flow}.

In fluid dynamics-related contexts \citep{vignonRecentAdvancesApplying2023}, DRL typically involves the interaction of an agent with a physical environment, as illustrated in Figure \ref{fig:drl}. The agent, implemented as a neural network, begins with no prior knowledge of the underlying physical phenomena. Using on-policy algorithms like Proximal Policy Optimization (PPO) \citep{schulmanProximalPolicyOptimization2017}, the agent receives an observation—a partial and potentially noisy representation of the environment’s full state. The state represents the complete set of information describing the environment, but is often not fully accessible to the agent. The observation, which is a subset of this state, is used as input to the policy. The agent then receives a reward, a scalar value that indicates how favourable the observation is towards achieving the desired outcome. Based on the observation and the reward, the agent selects an action (or set of actions), which modifies the environment, leading to a new state in the next step.

When the environment (e.g. the DNS of a turbulent flow) receives a new set of actions, the physics of the flow are simulated to produce a new observation (e.g. a sample of the
velocity field) and its associated reward. This observation updates the environment's state, and the agent uses the collected experience to adjust the neural network's weights based on the current reward and the chosen policy. These updated weights improve the agent's decision-making process for future actions. The agent's new actions then interact with the environment, leading to another state transition, which triggers the next step. Multiple steps form an episode, and the agent continuously collects experiences over many episodes, using an explicit optimisation step to update the policy and maximise cumulative rewards.

In this work, an agent is trained using the PPO algorithm \citep{schulmanProximalPolicyOptimization2017} to gather observations from a Direct Numerical Simulation (DNS) (\textit{the environment}) of a plane, turbulent channel flow and apply a time-dependent \textit{action} %$\omega$ 
that dynamically modifies the wall boundary conditions, ultimately pursuing a \textit{policy} that maximises a given the reward.

We will be looking at two different rewards: one is related to the minimisation of  $C_f$, the other to the promotion of highly elongated velocity streaks.

This second choice is motivated by the intention of understanding how the topology of coherent structures impacts the development of the wall regeneration cycle. This involves promoting specific topologies of the flow, which helps infer the causality between the control law and its effect on the turbulent wall-cycle by linking the straightening of streaks to the reduction of $C_f$.
Further efforts will focus on this direction by testing alternative objective functions that target specific features of the wall cycle.

From an implementation perspective, this work involved integrating the \textit{StableBaselines3} DRL libraries \citep{JMLR:v22:20-64} with the open-source DNS solver \textit{CaNS} \citep{costaFFTbasedFinitedifferenceSolver2018}. The final interface is based on an ad-hoc MPI implementation, where a Python code invokes the execution of the DNS code on a set of processors and manages the communication with the DRL library. While the most common choice is resorting to Python codes both for the DNS and the DRL routines (see for example \cite{vignonEffectiveControlTwodimensional2023}), similar unconventional couplings become necessary when the 3D turbulent DNS requires more tailored codes not to waste computational resources, like in the work of \cite{guastoniDeepReinforcementLearning2023}, \cite{suarezFlowControlThreedimensional2024} and \cite{fontActiveFlowControl2024}. Alternatives that are simpler to implement have been explored, such as DNS codes being asked to save data on disk, rather than keeping it in memory, with bash scripts orchestrating the various calls of the DRL training and the DNS code to produce new data. This method has been initially implemented with a reliable DNS in-house developed solver, SUSA \citep{montiGenesisDifferentRegimes2020}\citep{montiLargeeddySimulationOpenchannel2019}, but it has later been discontinued, due to the high inefficiency of the coupling itself.

The initial results indicate the need for more advanced control laws to fully exploit the tools developed in these early experiments. Currently, DRL agents are trained to find a control law based on a single dynamic parameter, but better policies could emerge from more complex control laws involving multiple dynamic parameters.

\section{Methodology}\label{sec2}
\subsection{Deep Reinforcement Learning}

\begin{figure}[h!]
	\centering
	\input{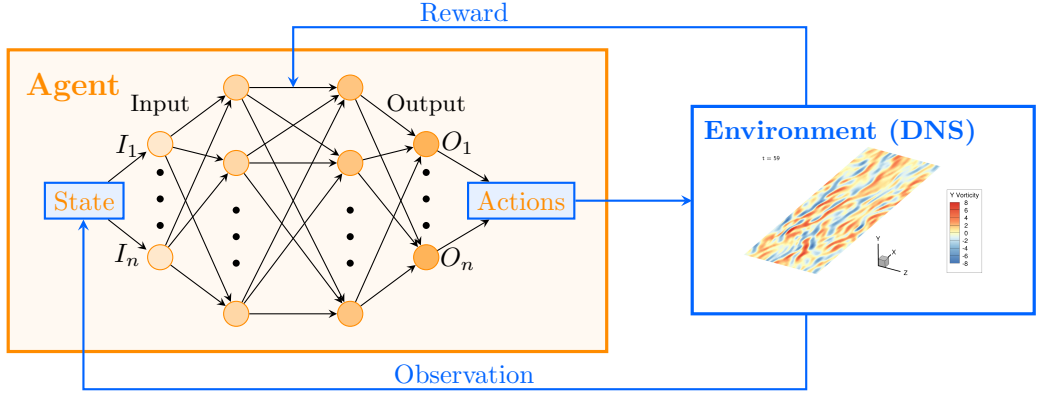}
	\caption{Schematic representation of the interaction between the DRL agent and the DNS environment.}
	\label{fig:drl}
\end{figure}
As mentioned, in the present context, DRL-based flow control consists 
in the interaction of an agent with a turbulent channel flow modeled through a DNS, as depicted in figure \ref{fig:drl}. 

The agent aims to learn a policy $ \pi_\theta(a|s) $ that represents the probability of taking action $ a $ in state $ s $, given neural network parameters $\theta $. The interaction with the environment generates a trajectory $ \{(s_t, a_t, r_t, s_{t+1})\} $ in the state-action space based on the current policy. The agent then evaluates the advantage function $ A_t $, which measures how much better a particular action is compared to the average action taken in a given state. 
$A_t $ is typically defined as:
\begin{equation}
A_t = Q(s_t, a_t) - V(s_t)
\end{equation}
where $ Q(s_t, a_t)$ is the action-value function, and $ V(s_t)$ is the value function.

However, in Proximal Policy Optimization (PPO), the agent does not have access to the true values of $ Q(s_t, a_t)$ or $ V(s_t)$. Instead, PPO approximates the advantage function using the critic network (value function network) and the sampled returns from the environment. Specifically, the advantage function is approximated by:
\begin{equation}
A_t \approx R_t - V(s_t)
\end{equation}
where $ R_t $ is the sampled return, and $ V(s_t) $ is the value function estimated by the critic.

The policy parameters $ \theta$ are then updated to minimise the following loss function:
\begin{equation}
L(\theta) = \mathbb{E}_t \left[ L^{CLIP}(\theta) - c_1 L^{VF}(\theta) + c_2 S[\pi_\theta](s_t) \right],
\end{equation}
where
\begin{equation}
L^{CLIP}(\theta) = \mathbb{E}_t \left[ \min \left( r_t(\theta) A_t, \text{clip}(p_t(\theta), 1 - \epsilon, 1 + \epsilon) A_t \right) \right],
\end{equation}
and
\begin{equation}
L^{VF}(\theta) = \mathbb{E}_t \left[ \left( V_\theta(s_t) - V_t^{\text{target}} \right)^2 \right].
\end{equation}

Here, $ p_t(\theta) = \frac{\pi_\theta(a_t|s_t)}{\pi_{\theta_{old}}(a_t|s_t)} $ is the probability ratio, and $ \epsilon$ is a hyperparameter controlling the clipping range. $ V_t^{\text{target}}$ is typically computed as the discounted sum of rewards plus the bootstrapped value of the next state. The terms $ c_1$ and $ c_2 $ are coefficients that balance the contributions of the value function loss and the entropy bonus, respectively. The entropy bonus is defined as:
\begin{equation}
S[\pi](s) = - \sum_a \pi(a|s) \log \pi(a|s),
\end{equation}
encouraging exploration by discouraging premature convergence of the policy.

The policy outputs a probability distribution over possible actions, and during training, an action $a$ is sampled from this distribution based on the predicted probabilities.

The input data from DNS can have any structure, as long as the neural network (NN) is built with matrices of compatible sizes for multiplication. Inputs can be array-like, and in such cases, the most common policy adopted is the Multi-Input Policy (MLP) \citep{lillicrapContinuousControlDeep2019}, which learns to weigh each input based on its relevance to the task, but it does not explicitly model any interdependencies between input features. However, in our specific application, it is important to preserve the spatial correlation of the input, as any combinations of sampled instantaneous quantities from a flow field are linked by the Navier-Stokes equations. Therefore, 2D, black and white, images rather than one-dimensional arrays of re-arranged data are fed to the neural network. This data consists of wall-parallel slices of streamwise velocity values ($96 \times 64$) sampled at a given $y^+$ (i.e. $y^+=20$). When using an image as input, the colour of a given pixel is not necessarily uncorrelated with the colours of other pixels.

By using images, we can effectively exploit the features of convolutional neural networks (CNNs) \citep{lecunBackpropagationAppliedHandwritten1989}\citep{lecunGradientbasedLearningApplied1998} without losing information. These policies could show a computational advantage to more standard fully connected layers, reducing computational requirements that otherwise could become prohibitive, and they are effective in fluid dynamics for inspecting turbulent fields as images. This approach is illustrated in the work of \cite{guastoniConvolutionalnetworkModelsPredict2021}, where DRL with CNNs was applied to another flow control technique, opposition control, achieving notable results in terms of drag reduction. A more schematic but detailed representation of how the training and the interaction with the DNS environment take place can be found in algorithm \ref{alg:step}.

The custom CNN built for this purpose begins by applying a series of convolutional layers, which progressively learn to extract increasingly complex features from the input. In the initial stage, the network employs a convolutional layer with 48 filters, each using a 3x3 kernel. This layer scans the input, detecting basic patterns like edges and textures. After each convolution, a batch normalisation step stabilises the training process by normalising the output, and a ReLU (Rectified Linear Unit) activation function introduces non-linearity, enabling the network to learn more complex patterns.

As the data progresses through the network, it encounters additional convolutional layers, first with 64 filters and then with 96 filters, both using the same 3x3 kernel size. More layers will lead to more complete but simplified representation of turbulent snapshots corresponding to the observed state. Each convolutional layer is again followed by batch normalization and a ReLU activation to ensure that the learning process remains robust and capable of capturing complex features.

Following the convolutional layers, the network transitions to fully connected layers. First, the high-dimensional output from the convolutional layers is flattened into a one-dimensional vector. This vector, now representing a combination of all the features extracted, is passed through a series of fully connected layers. The first of these layers reduces the vector to 96 units, and the second further reduces it to 64 units, both using ReLU activations to retain the network’s ability to model complex relationships within the data.

Finally, the output of the last fully connected layer is passed through another linear layer, which reduces the feature vector to the desired output size specified by the user. 
The dimension of this output, which subsequently becomes the input for both the actor and critic networks, has been determined through empirical tests within the framework of this work and is the outcome of a trade-off argument. A value that is too small may fail to adequately represent complex data structures, while a larger dimension, although offering a better representation of turbulent fields, increases memory usage and extends learning time. This can slow down training to the extent that useful policies may not be learned within the given network architecture.

As far as the critic/value network is concerned, two hidden layers of 512 neurons each are used for both. Smaller sizes have been tried (256, 384) but they struggled to deal with the dimensionality of the input or with the number of features to be extracted (always 16).

\begin{figure}[h!]
	\centering
	\includegraphics[width=1\linewidth]{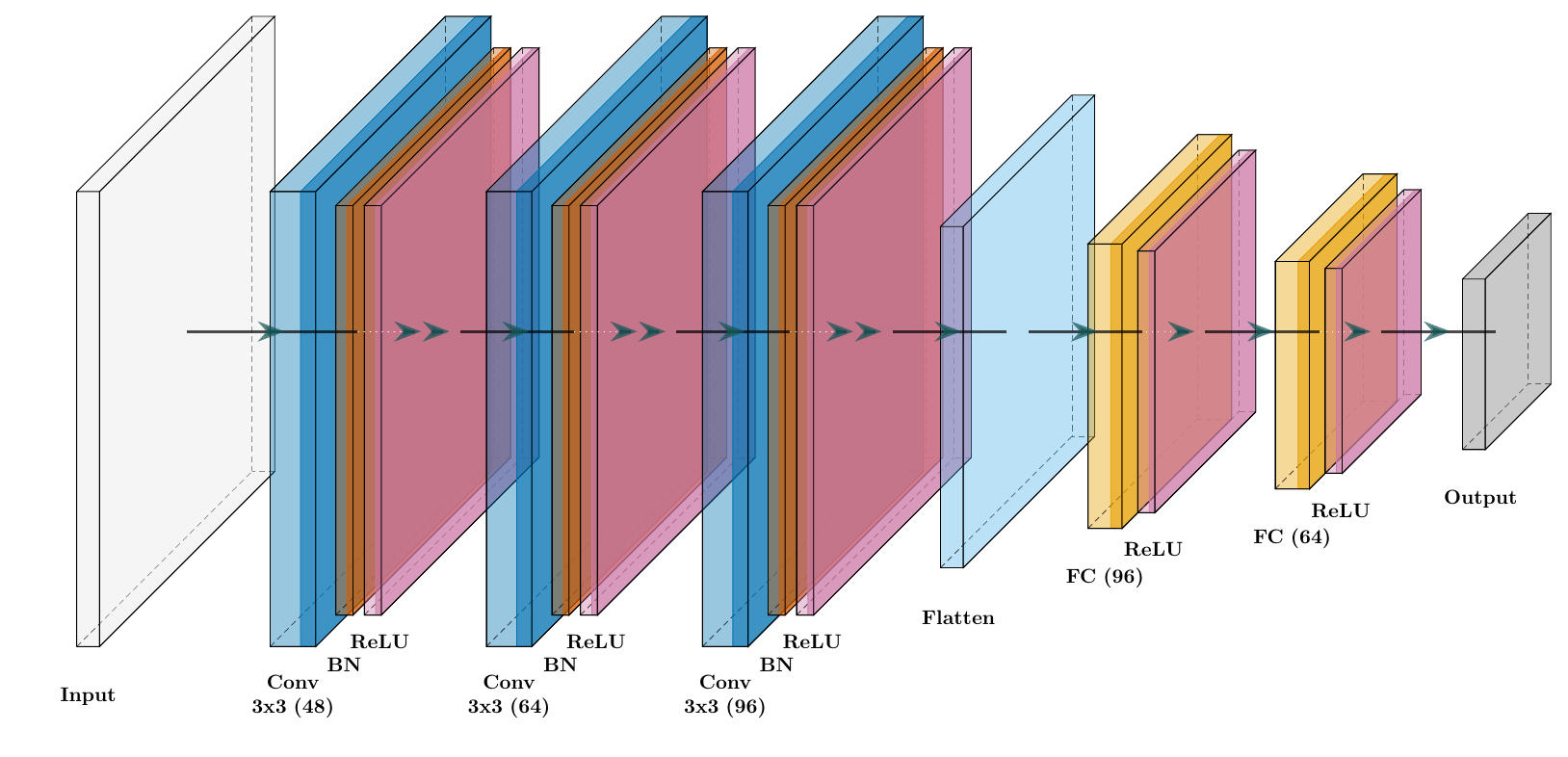}
	\caption{Schematic representation of the CNN neural network. The dimensions are indicated in brackets, ``BN'' stands for Batch Normalisation and ``FC'' for Fully Connected layers.}
	\label{fig:network}
\end{figure}

\subsection{\textcolor{mygreen2}{DNS set up}}
As alrea\text{d}y mentioned, our \textit{environment} is the DNS of a turbulent plane channel flow with modified wall boundary conditions. Thus, we solve the full 3D, unstea\text{d}y Navier-Stokes (NS) equations,
in their incompressible formulation:

\begin{align}
\pdv{u_i}{t} + u_j\pdv{u_i}{x_j} = -\frac{1}{\rho}\pdv{p}{x_i} + \nu\pdv{u_i}{x_j}{x_j} \label{eq:ns1}\\
\pdv{u_i}{x_i} = 0,
\end{align}
where the Einstein index notation, with summation over repeated indexes, is used to represent the differential operators, $x_{\{i,j,k\}}$ with $i = \left[1,2,3\right]$ corresponds to the $i^{\text{th}}$ direction in a three-dimensional space, $u_i$ is the corresponding velocity component, $p$ is the pressure, $\rho$ the density and $\nu$ the kinematic viscosity of the flow. 

To advance the equations in time we use two different solvers,
\textit{CaNS} \citep{costaFFTbasedFinitedifferenceSolver2018} and SUSA \citep{montiGenesisDifferentRegimes2020}\citep{montiLargeeddySimulationOpenchannel2019}. They are both efficient solver for massively-parallel direct numerical simulations of incompressible turbulent flows.

\textit{CaNS} is a freely available code \citep{GitHubCosta} designed for simulating parallel turbulent flows. The code uses a second-order, staggered Cartesian finite grid pressure-correction scheme \citep{chorinNumericalSolutionNavierStokes1968}, advanced in time through a three-stage, low-memory Runge-Kutta scheme \citep{rai1991direct}. Additionally, the code offers an optional implicit treatment of viscous terms in the wall-normal direction. The pressure Poisson equation, solved at each Runge-Kutta stage, is handled using eigenfunction expansions, allowing the use of highly efficient FFT-based solvers \citep{cooleyAlgorithmMachineCalculation1965} for problems with various combinations of homogeneous pressure boundary conditions. \textit{CaNS} incorporates a 2D pencil-like domain decomposition, enabling efficient massively-parallel simulations with excellent strong-scaling performance across thousands of cores. Extensive validation cases are available in the code repository \citep{GitHubCosta}, including turbulent Poiseuille flow, turbulent Couette flow, and turbulent square duct flow. The code supports Dirichlet or Neumann boundary conditions in one direction, periodic conditions in another, and a combination of Dirichlet, Neumann, or periodic conditions in the third direction.

\textit{SUSA} is a second-order finite volume code (in both time and space) developed at City St George's, University of London. It has been extensively validated for both DNS and LES versions in a wide range of turbulent flows, with a full list available in \citep{montiGenesisDifferentRegimes2020} and the references therein. The code supports a periodic direction (spanwise), while in the wall-normal plane, it adopts a body-fitted approach compatible with Cartesian, C, and O grids. \textit{SUSA} uses a discretisation of the momentum and mass conservation equations on a cell-centered, co-located grid. Fluxes are approximated using a second-order central formulation, and the Rhie and Chow method \citep{rhie1983numerical} is employed to prevent spurious pressure oscillations. The equations are advanced in time through a second-order semi-implicit fractional-step procedure, where the implicit Crank-Nicolson scheme is applied for wall-normal diffusive terms, and the explicit Adams-Bashforth scheme is used for all other terms. Further details can be found in \citep{montiGenesisDifferentRegimes2020} and the references therein.

\begin{figure}[h!]
	\centering
		\begin{tikzpicture}[x={(1,-.25,-.25)}, y={(0,1,0)}, z={(.25,0,-1.25)}, scale=0.5,
		%Option for nice arrows
		>=stealth, %
		inner sep=0pt, outer sep=2pt,%
		axis/.style={thick,->},
		wave/.style={thick,color=#1,smooth},
		]
		
		% Colors
		\colorlet{darkgreen}{green!50!black}
		\colorlet{lightgreen}{green!80!black}
		\colorlet{darkred}{red!50!black}
		\colorlet{lightred}{red!80!black}
		
		%%% Define quantities
		\pgfmathsetmacro{\step}{3.14/3} % to plot sttw
		\pgfmathsetmacro{\steparrow}{3.14/8} % to plot sin arrows
		\pgfmathsetmacro{\ends}{3*3.14} % to plot sttw
		
		\pgfmathsetmacro{\lx}{3.14} % axis length
		\pgfmathsetmacro{\ly}{3.14} % axis length
		\pgfmathsetmacro{\lz}{3*3.14} % axis length
		
		\pgfmathsetmacro{\platethick}{3.14/18} % plate thickness
		
		\pgfmathsetmacro{\amp}{1} % wave amplitude

		% ------- PLOT --------
		\coordinate (O) at (0,0,0); % to start with
		\coordinate (OO) at (-\lx*1.5,0,-\lx); % to start with
		
		% Lower plate
		\path[draw=black, line width=0.5pt, line width=0.4pt, top color=gray!35,bottom color=gray!65,middle color=gray!50,shading angle=-95] (-\lx,0,0) -- (\lx,0,0) -- (\lx,-\platethick,0) -- (-\lx,-\platethick,0) -- cycle; % front
		
		\path[draw=black, line width=0.5pt, line width=0.4pt, top color=gray!15,bottom color=gray!55,middle color=gray!20,shading angle=-140] (\lx,-\platethick,0) -- (\lx,-\platethick,\lz) -- (\lx,0,\lz) -- (\lx,0,0) -- cycle; %side
		
		\path[draw=black, line width=0.5pt, line width=0.4pt, top color=gray!5,bottom color=gray!55,middle color=gray!10,shading angle=-140] (-\lx,0,0) -- (-\lx,0,\lz) -- (\lx,0,\lz) -- (\lx,0,0) -- cycle; %top 

		% Frame
		\draw[axis] (OO) -- +(\lx/2+\step,0,0) node [right] {$z$};
		\draw[axis] (OO) -- +(0,\ly/2,0) node [above] {$y$};
		\draw[axis] (OO) -- +(0,0,\lz/8+\step) node [right] {$x$};
		
		% StTW
		\draw[wave=myblue, variable=\z,samples at={0,\step,...,\ends}]
		plot ({\amp*sin(2*\z r)},0,\z);
		\draw[myblue, line width=0.2pt] (0,0,0) -- (0,0,\lz);
		
		% StTW arrows
		\foreach \z in{0,\steparrow,...,\ends}
		\draw[color=myblue,-stealth] (0,0,\z) -- ({0.9*\amp*sin(2*\z r)},0,\z);
		
		% Upper plate
		\path[draw=black, line width=0.5pt, line width=0.4pt, top color=gray!35,bottom color=gray!65,middle color=gray!50,shading angle=-95] (-\lx,0+\ly,0) -- (\lx,0+\ly,0) -- (\lx,-\platethick+\ly,0) -- (-\lx,-\platethick+\ly,0) -- cycle; % front
		
		\path[draw=black, line width=0.5pt, line width=0.4pt, top color=gray!5,bottom color=gray!20,middle color=gray!10,shading angle=-140] (\lx,-\platethick+\ly,0) -- (\lx,-\platethick+\ly,\lz) -- (\lx,0+\ly,\lz) -- (\lx,0+\ly,0) -- cycle; %side
		
		\path[draw=black, line width=0.5pt, line width=0.4pt, top color=gray!5,bottom color=gray!35,middle color=gray!10,shading angle=-140] (-\lx,0+\ly,0) -- (-\lx,0+\ly,\lz) -- (\lx,0+\ly,\lz) -- (\lx,0+\ly,0) -- cycle; %top 
		
		% StTW - up
		\draw[wave=myblue, variable=\z,samples at={0,\step,...,\ends}]
		plot ({\amp*sin(2*\z r)},\ly,\z);
		\draw[myblue, line width=0.2pt] (0,\ly,0) -- (0,\ly,\lz);
		
		% StTW arrows - up
		\foreach \z in{0,\steparrow,...,\ends}
		\draw[color=myblue,-stealth] (0,\ly,\z) -- ({0.9*\amp*sin(2*\z r)},\ly,\z);
		
		% Lambda_z
		\draw[<->, mygreen, line width=0.5] (-\lx/2.5,\ly,\ends/4) -- (-\lx/2.5,\ly,\ends/3+\ends/4) node[xshift=-22pt,yshift=-3pt]{$\kappa_x$};
		
		% sin(kx - omegat)
		\node[myblue] at (-\lx/2,\ly,+\lz/9*10.5) {$w = A \, \sin \left(\kappa_x x - \omega\left(t\right) t \right)$};
		
		% Heights
		\draw[black!80, thin, dashed] (-\lx,\platethick,0) -- (-\lx,-\platethick+\ly,0); %left
		\draw[black!80, thin, dashed] (\lx,\platethick,0) -- (\lx,-\platethick+\ly,0); %center
		\draw[black!80, thin, dashed] (\lx,\platethick,\lz) -- (\lx,-\platethick+\ly,\lz); %right
		\draw[stealth-stealth] (\lx*1.05,0,\lz) -- (\lx*1.05,-\platethick+\ly,\lz) node[xshift=10pt,yshift=-20pt]{$2H$}; %arrow
		
		% Lengths
		\draw[stealth-stealth] (-\lx*1.1,\ly,0) -- (-\lx*1.1,\ly,\lz) node[xshift=-65pt,yshift=-28pt]{$L_x$}; %x
		\draw[stealth-stealth] (-\lx,0,-\lx*0.2) -- (\lx,0,-\lx*0.2) node[xshift=-20pt,yshift=-6pt]{$L_z$}; %z
		
		% Mean flow
		\draw[->, myorange, line width=3] (-\lx/4,\ly/3,-\ends/2.6) -- (-\lx/4,\ly/3,\ends/12) node[xshift=-28,yshift=-30pt]{\textbf{$U_b$}};

	\end{tikzpicture}
	\caption{Channel flow DNS setup. The STW boundary condition is applied equally at the two walls. Now $\omega$ is a function of $t$, so the phase-speed is not constant any more. This is still a class of Dirichlet BCs, that along with homogeneous Dirichlet for $u$ and $v$ and no-slip conditions are applied at the two walls. Transparent sides in this figure correspond to periodic boundary condition, in the streamwise and spanwise directions.}
	\label{fig:chan}
\end{figure}
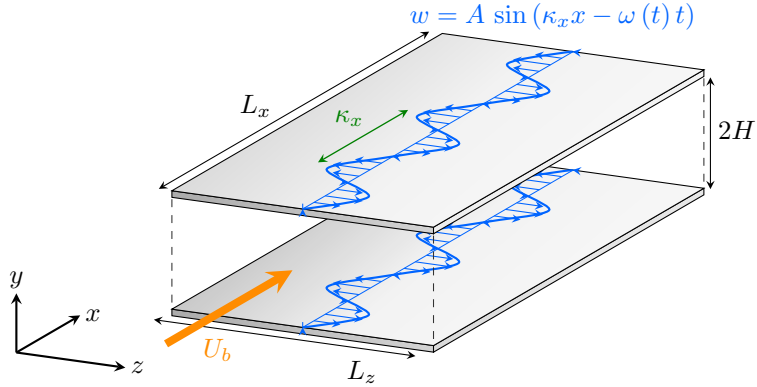

The baseline codes have been modified to account for time-dependent boundary conditions at the walls. Specifically, we have implemented modified conditions on the spanwise velocity components $w$ to introduce the effect of the streamwise travelling waves at the two walls of a fully turbulent channel flow. The boundary condition on $w$ reads as:
\begin{equation}
w\left(x,0,z,t\right) = A \sin \left(\kappa_x x - \omega t \right),
\label{eq:wt}
\end{equation}
where $\omega$ is a pulsation, $A$ is an amplitude, and $\kappa_x$ is a spatial wavenumber. The same conditions are applied at $y=0$ and \(y=2H\), where \(H\) is the semi-height of the channel, as shown in figure \ref{fig:chan}.
%       $\left(n_x, n_y, n_z\right)$ & $\left(96, 100, 64\right)$ & $\left(96, 100, 64\right)$ \\
  %      $\left(L^+_x, L^+_y, L^+_z\right)$ & $\left(628, 200, 314\right)$ & $\left(628, 200, 314\right)$ \\
  %      $Re_b$, \textcolor{myblue}{$Re_{\tau}$} \textcolor{myblue}{(without STW)} & 6340, \textcolor{myblue}{$200$} & 6340, \textcolor{myblue}{$200$} \\

\begin{figure}[h!]
	\centering
	\includegraphics[width=0.75\linewidth]{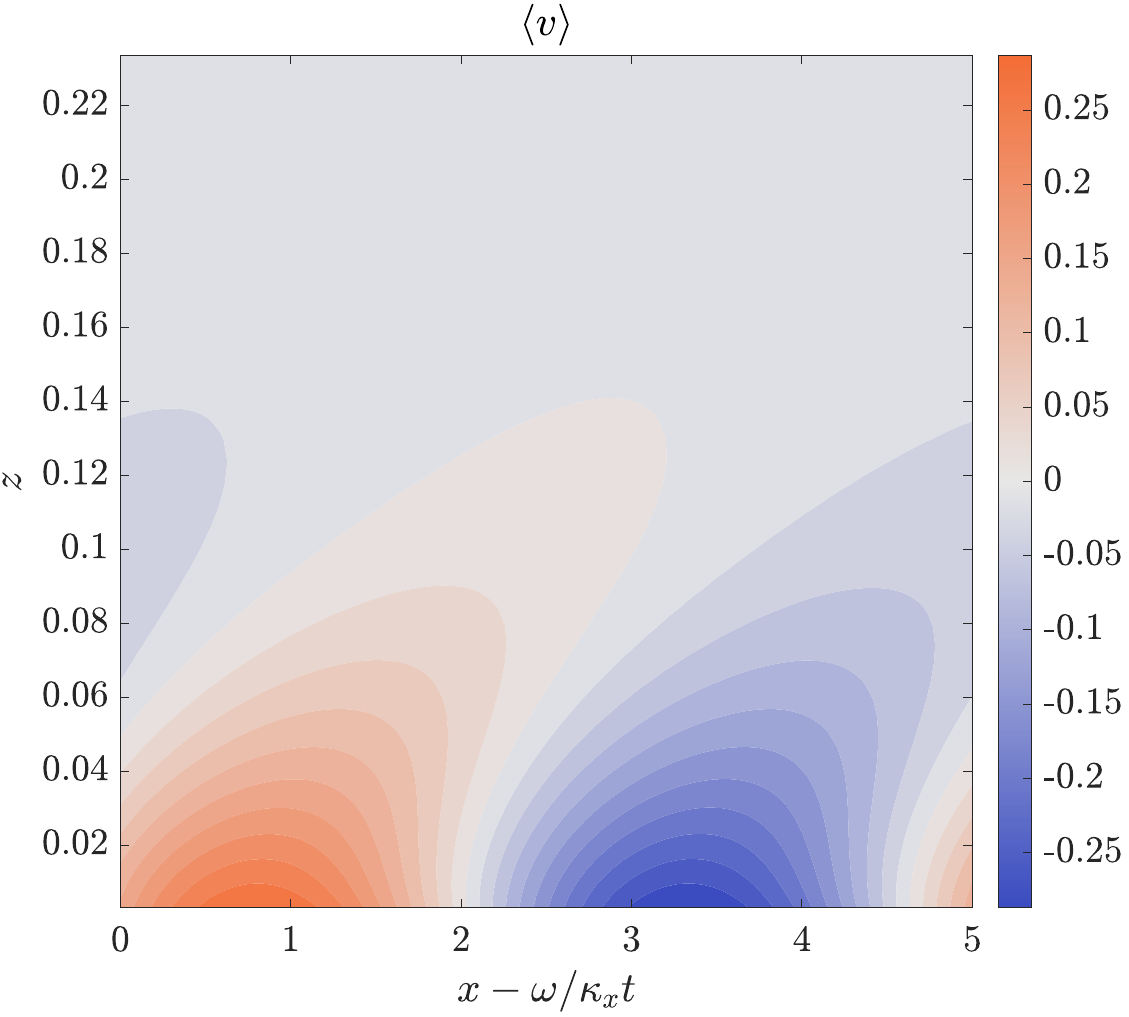}
	\caption{Average of the spanwise velocity in phase with the wave in a region close to the wall. The values of $\langle v \rangle$ are reported in the colorbar. The vertical represents the wall-normal coordinate in outer units; the horizontal axis corresponds to the convective variable $\xi$ in phase with the wave. The positive peak and its negative counterpart are well represented at the wall.}
	\label{fig:vv_phase}
\end{figure}

\begin{figure}[h!]
	\centering
	\includegraphics[width=1\linewidth]{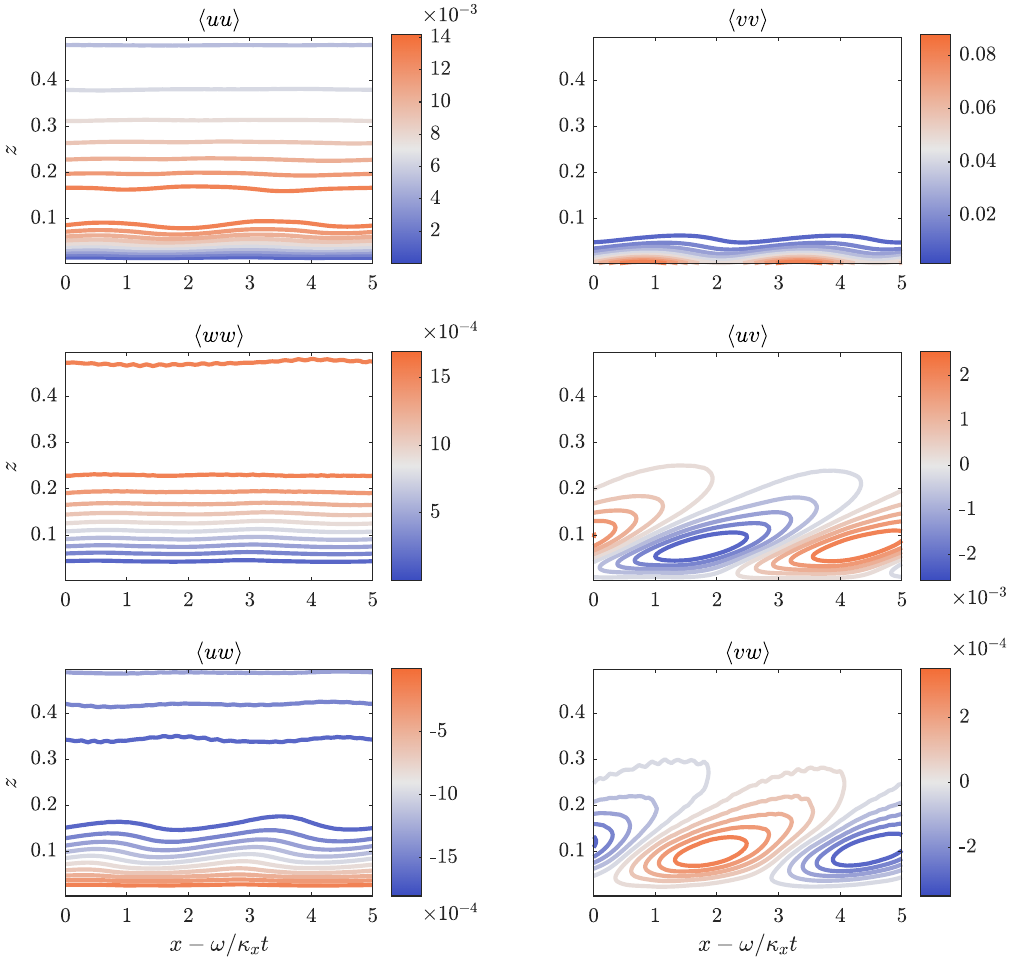}
	\caption{Reynolds stresses $\langle uu \rangle$, $\langle vv \rangle$, $\langle ww \rangle$, $\langle uv \rangle$, $\langle uw \rangle$ and $\langle vw \rangle$ in a region close to the wall. Values are in outer units to allow for a direct comparison with figure 10 of \citep{quadrioStreamwisetravellingWavesSpanwise2009}. The axes share the values of figure \ref{fig:vv_phase}, with different colorbars for each quantity. Reynolds stresses involving spanwise velocity fluctuations feature patterns correlated with the wall motion; the others present a thick layer with small, but still visible, footprints of the STW.}
	\label{fig:stresses_phase}
\end{figure}

All simulations have been performed at the same $Re_b=U_b h /\nu=3170$ (where $U_b$ is the bulk velocity, $h$ is the semi-height of the channel and $\nu$ is the kinematic viscosity) which corresponds to $Re_{\tau}=\frac{u_{\tau} h}{\nu} =200$ (where $u_{\tau}=\sqrt{\tau_w/\rho}$ is the friction velocity, $\left. \tau_w=\mu \frac{dU}{dy} \right|_{y=0}$ being the wall-shear stress), using the same domain size (streamwise dimension $L_x/h=\pi$, spanwise dimension $L_z/h=\pi/3$), and the same number of grid nodes ($96 \times 100 \times 64$,  along the streamwise $x$ direction, the wall-normal $y$ direction and the spanwise $z$ direction, respectively). Before considering DRL-actuated cases, both codes have been validated, using the aforementioned parameters, for both non-actuaded channel flows and for an actuated STW case \citep{quadrioStreamwisetravellingWavesSpanwise2009} with $\kappa_x^+ = 0.0063$ and $\omega^+ = 0.0195$ \citep{gattiReynoldsnumberDependenceTurbulent2016} (units with a $+$ superscript are normalised with $\nu$ and $u_{\tau}$). In figures \ref{fig:vv_phase} and \ref{fig:stresses_phase}, we report the spanwise velocity and the diagonal Reynolds stresses distribution for the STW actuated channel in the case of a drag reducing setup. The obtained results are in good agreement with the ones in \citep{quadrioStreamwisetravellingWavesSpanwise2009} (see figures 9 and 10 within the given reference). In the given figures, the averages are obtained in phase with the wall motion, i.e. computed accumulating data at $\xi = x - \omega/\kappa_x t$. Those averages, denoted by $\langle \cdot \rangle$, are computed using a 2D average over the spanwise direction and time. In particular, statistics have been accumulated over $\approx 150$ periods of the STW after the initial numerical transient.

As far as all the DRL-controlled cases reported, the nominal value of $Re_{\tau} = 200$ (computed in the corresponding non-actuated case) results in a domain size, expressed in viscous units, of $L^+_x=628$, $L^+_y=200$, and $L^+_z=314$. These values were selected for several reasons: they are slightly larger than the minimum requirements for a self-sustained \textit{minimal flow unit} channel flow \citep{jimenezMinimalFlowUnit1991}; they provide a numerical domain that accommodates at most two pairs of velocity streaks, simplifying the analysis of the effects of actuations on coherent structures; and they allow for fast evaluation of the environment step without imposing significant memory usage requirements. Regarding the grid size, $\Delta x^+$ and $\Delta z^+$ are well within the commonly recommended values \citep{moser1999direct}, and the distance of the first internal grid point from the wall is consistently below $y^+=1$ for all considered cases.

\subsection{Actuation and objective functions}

The connection with DRL is embedded in \(\omega\), which becomes a time-dependent parameter (i.e., \textit{the actuation}), whose behaviour now depends on time and varies according to the estimations of the \textit{agent} to achieve a better reward. Most of the work carried out so far has focused on using DRL to find a time-law \(\omega(t)\) to maximise a chosen objective function during an \textit{episode}.

It is noted that using \(\omega(t)\) as the actuation or the actual phase \(\phi(t) = \omega(t) \cdot t\) in Eq.~\ref{eq:wt} represents an equivalent choice, as both approaches describe the same oscillatory behavior.

To justify the choice of \(\omega\) as an \textit{actuator}, it is necessary to provide some background on the physical modifications that STWs introduce in the near-wall region. The following conceptual description closely follows the discussion offered by \citep{quadrioStreamwisetravellingWavesSpanwise2009}. The mechanism induced by STWs involves two main effects: phase locking between the traveling waves and near-wall turbulent structures, and the interaction of the turbulent flow with the Stokes layer generated by the oscillatory, time-dependent condition on the spanwise velocity at the wall.

When the phase speed of the traveling waves, \(c = \omega / k_x \), approaches the \textit{convective velocity} of the near-wall turbulent structures, \(\mathcal{U}_w\) (where \(\mathcal{U}_w \simeq 0.5 U_b\), \citep{kim1993propagation}), the near-wall flow structures synchronise with the wall waves. For STW with a fixed \(\omega\), as in \citep{quadrioStreamwisetravellingWavesSpanwise2009}, this phase-locking mechanism leads to a maximum drag increase when \(c = \mathcal{U}_w \simeq 0.5 U_b\). In our case, by allowing \(\omega\) to vary in time, it is possible to adapt to the local convection speed of the structures, since \(\mathcal{U}_w = 0.5 U_b\) is an average value representative of various local convection speeds.

Concerning the second mechanism, the traveling wave condition can be compared to the oscillating wall mechanism (i.e., the STW limit for \(k_x = 0\)). In the oscillating wall case, for small Stokes layer thicknesses (i.e., high frequencies), only a modest effect is exerted on the structures. A negligible effect also occurs for extremely low frequencies when the Stokes layer thickness exceeds the buffer layer size. Between these two extremes, there exists an optimal frequency \(\omega_{\text{opt}}\) that maximises drag reduction by weakening the regeneration cycle, thus reducing turbulence intensities. This optimal value of $\omega$ (\(\omega_{\text{opt}}\)), is related to the typical lifetime of coherent structures in the buffer layer. Similarly, for traveling waves, the flow embeds a Stokes layer, and its thickness again plays a fundamental role in determining turbulence suppression and drag reduction.

The important difference for STWs is that the Stokes layer thickness depends not only on \(\omega\) but also on the wavelength \(k_x\). The maximum interaction, resulting in turbulence suppression and drag reduction, occurs when \(|\mathcal{U}_w k_x - \omega|\) approaches \(\omega_{\text{opt}}\). In this case, \(\omega_{\text{opt}}\) is affected by how the traveling wave modifies the expected lifetime of the coherent structures (e.g., structures phase-locked with the traveling wave tend to survive longer). In the present case, when considering a minimal flow unit, it is expected that by varying \(\omega\) as a function of time, the Stokes layer thickness can be optimised (\textit{modulated}) based on the actual flow configuration.

Another parameter that needs to be considered is the duration of an episode. To this end, one should consider the physical phenomenon at hand. In our particular case, it is known that STWs can affect the presence/position of the low-speed streaks in the logarithmic region close to the wall. The streaks are characterised by a life cycle of \(\Delta t^+_{\text{cycle}} = 80-100\) (see \cite{jimenezAutonomousCycleNearwall1999}), thus an episode length \(t_{\text{ep}}^+ = 160\) is suitable to capture, with some margin of uncertainty, at least one full regeneration cycle. We have also considered some cases with \(t_{\text{ep}}^+ = 80\), finding that this time interval is not always sufficient to control a full cycle, and other cases with a larger time interval, i.e., \(t_{\text{ep}}^+ = 320\) that did not present any sigificant variation in the results. 

Within the duration of each episode, \(\omega\) is set to change every \(\Delta \tau\), assigning a smoothed piece-wise constant behaviour to avoid sharp changes in the boundary values across the actuations. Note that \(\Delta \tau\) is the actuation period (also called a \textit{step}), and its duration was initially set to 10 viscous time units to ensure a sufficient number of changes during an episode while not excessively interfering with the actuations. It was found that a smaller \(\Delta \tau\) had less impact on the outer flow and led to unwanted oscillations in the flow statistics.

As previously discussed, the DRL agent needs to be trained to learn how to choose the next action and achieve an effective policy to optimise the reward \(r\). The initial and most straightforward reward \(r\) for this problem is based on \(C_f\):
\begin{equation}
r = \frac{C_{f,\text{max}} - C_{f}}{C_{f,\text{max}} - C_{f,\text{min}}}
\label{reward1}
\end{equation}
In the above expression, \(C_{f,\text{max}}\) and \(C_{f,\text{min}}\) are, respectively, the maximum \(C_f\) that is considered (the one corresponding to an unactuated turbulent channel flow with no control at the wall) and the minimum one that is aimed for, corresponding to $65\%$ of DR, an optimistic goal based on results achieved with some other DR techniques coupled with DRL, such as by Guastoni et al. \citep{guastoniDeepReinforcementLearning2023}. This initial training with the $C_f-$based reward was carried out with the DNS-DRL coupling SUSA-Tensorforce via bash scripts (see details in section \ref{sec:drldns}). Another characteristic aspect of this initial study is that the observed space consisted of a 2D wall-parallel field of the streamwise velocity sampled at $y^+=1$. The rationale behind this choice was to provide data to the agent that was strongly correlated to the quantity that defined the reward, given that $C_f$ is proportional to $\text{d}U/\text{d}y$ at the wall. For the next studies, this value of $y^+$ was changed to acquire snapshots containing more information of the whole state of the DNS simulation, relaxing the constraint of the relation with the reward.

These values can be tuned to assign a high reward to realistically achievable flow configurations, but the tuning procedure is not trivial: if \(C_{f,\text{min}}\) is too low, the network will not learn because all the policies lead to a low reward and are discarded, while if \(C_{f,\text{min}}\) is too high, the network will learn how to reach a configuration that does not significantly reduce the drag.
With this formulation, a small \(C_f\) implies that some drag reduction (DR) is obtained and that an associated temporal law for \(\omega(t)\)
 has been achieved.

Alongside equation \ref{reward1}, we have considered a second type of reward that tackles the actual topology of the coherent flow structures populating the near-wall region.
This choice is motivated by the aim of discovering the causality between the emergence and decay of coherent structures. We targeted a measure of the coherence of the velocity streaks as the cost function. In particular, following Doohan et al. \citep{doohanMinimalMultiscaleDynamics2021}, we used DRL to maximise the energy content of the streamwise velocity fluctuations within a box localised on the walls. This objective should be seen as an attempt to produce streaks that are more \textit{rectilinear} and resilient to instabilities.
In particular, the reward has been designed to maximise the ratio $\tilde{E}_{ks}/\tilde{E}_k$, where
\begin{align}
\tilde{E}_{ks} = \int_0^{L_x} \int_0^{\delta_y} \int_0^{L_z} u'^2 \text{d}x\text{d}y\text{d}z.
\label{eq:eks} \\
\tilde{E}_{k} = \int_0^{L_x} \int_0^{\delta_y} \int_0^{L_z} \left(u'^2 +v'^2 + w'^2 \right) \text{d}x\text{d}y\text{d}z.
\label{eq:ek}
\end{align}
In equations \ref{eq:eks} and \ref{eq:ek}, $u'$, $v'$ and $w'$ are the fluctuation of the streamwise, wall-normal and spanwise velocity respectively, $\delta_y$ is chosen to define a box whose height in friction units is set to $\delta_y^+=60$, a size that should be sufficient to host the wall velocity streaks, while not extending too much beyond the buffer zone. 
%\textcolor{myblue}{It is also worth noticing that the spanwise length of this box is enough to host at least a couple of counter-rotating streaks to let the DRL agent focus on a single, non-repetitive in the spanwise direction phenomenon. A bigger domain in the spanwise direction would simply allow more streaks to be hosted in the box, putting the DRL agent in the difficult position of controlling more streaks at the same time, which would definitely require more resources and wouldn't necessarily lead to more relevant considerations to be made concerning the re-generation mechanism of the single streaks. The domain size is still compliant with the minimal channel units \citep{jimenezMinimalFlowUnit1991}, so the streaks in the box are physically represented properly.}

Maximising the ratio
\begin{equation}
r = \tilde{E}_{ks}/\tilde{E}_k
\label{eq:rew2}
\end{equation}
should promote the longitudinal straightness of the streaks by promoting fluctuations in the streamwise direction. The rationale behind this experiment is related to the minimisation of the sinusoidal instability \citep{kawaharaLinearInstabilityCorrugated2003} of the streaks and its possible impact on the wall regeneration cycle.

One final matter to be discussed concerns the size of the DNS computational box. A large box would be computationally challenging for the training of the DRL network since it would involve a much larger dataset featuring complex and non-local flow interactions. This would lead to a long time required for completing the training. On the other hand, a box that is too small could lead to re-laminarisation of the flow for sizes below the minimal flow unit threshold, as indicated by Jim{\'e}nez and Moin in their 1991 seminal paper \citep{jimenezMinimalFlowUnit1991}. 

For this stu\text{d}y, a box size slightly larger than the minimal flow unit \citep{jimenezMinimalFlowUnit1991} was chosen. This choice is based on the hypothesis that a pair of velocity streaks contains all the required dynamics to sustain the wall regeneration cycle of turbulence, and that the DRL policy will be tailored to this condition.

\subsection{DRL-DNS interface}\label{sec:drldns}
Coupling the DNS solver with the DRL interface occupied most of the time needed to carry out this work. Initially, a finite volume method solver previously mentioned, SUSA, has been re-designed to interact with bash scripts that commanded when to write the files needed for the training, and to re-initialise the grid with the previous flow field with a new actuation parameter (if the new actuation belongs to the same episode). This is the setup that has been used, along with the Tensorforce libraries previously used by Vignon et al. \citep{vignonEffectiveControlTwodimensional2023}, for the cases where the reward was solely based on $C_f$, and the numerical setup for the episodes and the training is shown in table \ref{tab:cfd} in the left column.

\begin{table}[h!]
    \centering
    \begin{tabular}{lll}
        \toprule
        \textbf{Parameter} & \textbf{SUSA} & \textbf{CaNS} \\
        \midrule
        Incompressible DNS solver & SUSA & CaNS \\
        Simulation time $t^+$ & 80 & 160 \\
        Actuation time $\Delta \tau^+$ & 10 & 10 \\
        Cores per simulation & $32 + 1$ & $32 + 1$ \\
        $y^+$ for collecting data & 1 & 15 \\
        \midrule
        DRL agent policy & PPO & PPO \\
        DRL library & Tensorforce & StableBaselines3 \\
        Python-Fortran interface & bash & MPI (mpi4py) \\
        Input policy & Multi Input Policy & CNN \\
        \textcolor{myblue}{Actor/critic architecture} & [512,512], [512,512] & [512,512], [512,512] \\
        \bottomrule
    \end{tabular}
    \caption{Comparison of parameters used to train the network for reducing $C_f$ in the SUSA simulation and promoting straighter streak topologies in the CaNS simulation.}
    \label{tab:cfd}
\end{table}

The DRL portion of the code of the cases where the streaks topology is discussed relies on the use of \textit{StableBaselines3} library, an open source implementation of DRL based on \textit{PyTorch} \citep{paszkePyTorchImperativeStyle2019}. 
Despite not being optimal from a computational point of view, the use of Python is still the best option for a few reasons.
\begin{itemize}
    \item Strong community support: DRL is a machine learning methodology used in many different fields by both practitioners and developers. This vast base of users has produced many plug-and-play libraries available for public use, allowing, for example, engineers to interface their own software with sophisticated DRL libraries almost effortlessly. This enables researchers to focus on the aspects of the work closer to their area of expertise, easing and speeding up their research.
    \item Cross-platform libraries: The existence of Python environments that are easy to set up on different machines allows users to avoid spending too much time compiling/installing libraries.
    \item Libraries with Fortran/C++ bindings: The speed of compiled languages is still much higher than that of interpreted languages like Python. However, some Python-based packages such as NumPy \citep{harrisArrayProgrammingNumPy2020} embed popular, highly efficient, pre-compiled libraries, such as the popular C++ \textit{BLAS} and \textit{LAPACK} libraries. Others are just C++ wrappers, such as \textit{mpi4py} \citep{dalcinMpi4pyStatusUpdate2021}.
\end{itemize}

The overall software platform developed for the present work can be separated into two parts: the first part contains the external calls that define the duration of the network training and the settings for the agent; the second part concerns the environment, consisting of a class made up of several functions that are called during an actuation to obtain new observations and rewards, and to simulate the fluid flow.

The training is managed using a script that allows the user to change any settings. Additionally, a feature has been added that allows building a CNN from scratch, avoiding reliance on automated algorithms to set matrix sizes based on the input dimensions and the number of actions. In the current version, the customisation of the CNN parameters has been limited to the basic requirements needed to carry out a training session. In the future, other features that will add further flexibility will be implemented.

The development of a specialised interface is essential to balance computational efficiency in DNS simulations with the capabilities of state-of-the-art, open-source machine learning libraries. It is particularly noted that:
\begin{itemize}
    \item Compiled languages are more efficient in terms of speed and memory requirements than interpreted languages such as Python.
    \item Migrating a CFD code from one language to another, including the use of MPI-based calls, is a task that would require months, if not years. The effort would include programming, debugging, performance tuning, and extensive validation campaigns.
    \item A researcher with a background in turbulence and fluid dynamics may not have a sufficient skill set to implement efficient, state-of-the-art machine learning routines in a compiled language.
\end{itemize}

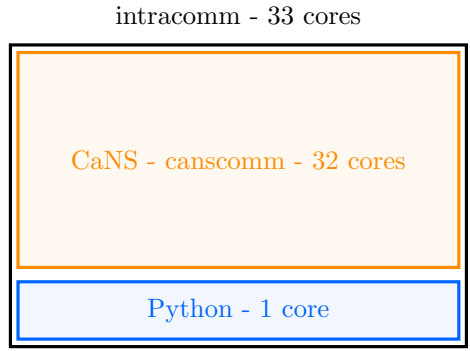
\begin{figure}[h]
\centering
\begin{tikzpicture}

 % Define dimensions
  \def\outerWidth{6}
  \def\outerHeight{4}
  \def\innerWidth{5.8}
  \def\innerHeight{3.8}
  
  % Draw the outer rectangle
  \draw[black, very thick] (0,0) rectangle (\outerWidth,\outerHeight);
  
  % Draw the first inner rectangle and add text
  \def\firstInnerX{0.1} % Starting X for the first inner rectangle
  \def\firstInnerY{0.1} % Starting Y for the first inner rectangle
  \draw[myblue, fill=myblue!5, very thick] (\firstInnerX,\firstInnerY) rectangle ++(\innerWidth,\innerHeight*0.2);
  \node at (\firstInnerX + 0.5*\innerWidth, \firstInnerY + 0.5*\innerHeight) {Text 1};
  
  % Draw the second inner rectangle and add text
  \def\secondInnerX{0.1} % Starting X for the second inner rectangle
  \def\secondInnerY{0.1+\innerHeight*0.25} % Starting Y for the second inner rectangle
  \draw[myorange, fill=myorange!5, very thick] (\secondInnerX,\secondInnerY) rectangle ++(\innerWidth,\innerHeight*0.75);

  % Text
  \node at (\secondInnerX + 0.5*\innerWidth, \secondInnerY + 0.5*\innerHeight*0.75) {\color{myorange}CaNS - canscomm - 32 cores};
  \node at (\firstInnerX + 0.5*\innerWidth, \firstInnerY + 0.5*\innerHeight*0.2) {\color{myblue} Python - 1 core};
  \node at (\secondInnerX + 0.5*\innerWidth, \outerHeight*1.1) {intracomm - 33 cores};

\end{tikzpicture}
\caption{Schematic of the MPI communicators for the DRL agent and CaNS with $N=32$. The autonomy of the CaNS environment is preserved with the existence of the \textit{canscomm} communicator, while \textit{intracomm} guarantees freedom of communication between the DNS ranks and the DRL one.}
\label{fig:mpi}
\end{figure}

These reasons constitute the rationale behind the development of our DNS-DRL software platform. In particular, we have followed a simple approach for linking
a Fortran/C++ DNS code to a Python DRL library. The basic idea concerns the execution of bash commands inside Python to launch the DNS environment for  data production.
 These data are used by the DRL agent that will update the weights of the neural network defining the policy. This approach is robust and easy to implement, but it is case-specific and cannot be generalised. For example, if the amount of data that the environment has to provide to the agent is too big, the time for the Fortran I/O interface to write the file on the disk and for the Python I/O interface to read it may become not negligible, and even saturate a node.
Moreover, for execution on large parallel architectures, the bash calls from Python must comply with the constraints of an HPC environment, such as requests to send a job in a shared queue and wait for its execution.
In the present work, these issues have been overcome using C-MPI wrappers for Python, with the \textit{mpi4py} package \citep{dalcinMpi4pyStatusUpdate2021}. Specifically, with this implementation, the main DRL code spawns \(N\) processes with the DNS executable, allowing a DNS to take place in an environment that is aware of the parent process by which it was spawned. This is formalised in the code by modifying the MPI initialisation in the DNS code, as shown by the example provided in figure \ref{fig:mpi}.
When the main code spawns the $N$ DNS parallel processes with \verb|mpi_comm_spawn| command, the DNS simulation is initialised with an MPI request to merge all the ranks and create an \textit{intracomm} communicator that contains all of the processes with 0 corresponding to the agent and $procs=1 \cdots N$ to the processes allocated to the DNS. Alongside the definition of a global set of processes, we also need to generate other two communicators, one encompassing all the DNS processes
(i.e. the, \textit{canscomm} communicator spanning $procs=0 \cdots N-1$)  and the other dedicated to the DRL agent.
Unfortunately, collective communications in the overall communicator (i.e. \textit{intracomm}) are not available yet within the \verb|mpi4py| library, so when the DNS solver has to share data with the DRL agent, \verb|mpi_gather| and \verb|mpi_gatherv| are used to collect data on rank 0 of \textit{canscomm} (which is also rank 1 of \textit{intracomm},) followed by a simple \verb|mpi_send| from rank 1 to 0 in \textit{intracomm} to make the data available for the Python library that manages the agent. 

Another important limitation of the actual \textit{mpi4py} library, concerns the \verb|mpi_comm_spawn| command that does not work on distributed resources, such as multiple nodes of an HPC environment. This sets a limit to the maximum number of ranks that can be used by the DNS code to $N-1$, being $N$ the number of cores available on a single node. This limitation is a serious handicap also for the maximum amount of memory that can be used.

From a more practical perspective, it should be mentioned that the training based on \textit{StableBaselines3} and \textit{CaNS}-DNS code is much more complex, in terms of subroutines and scripts to keep under control, than the one required for the previous implementation with SUSA and Tensorforce.
The main difficulty being related with the management of the communication between the environment and the ML libraries that now takes place using MPI calls and not I/O written on hard disks. Although further improvements are possible, the new implementation offers important advantages in terms of computational efficiency.

\section{Results}\label{sec3}
\begin{figure}[h!]
	\centering
	\begin{minipage}{.48\textwidth}
		\centering
			\includegraphics[width=1\linewidth]{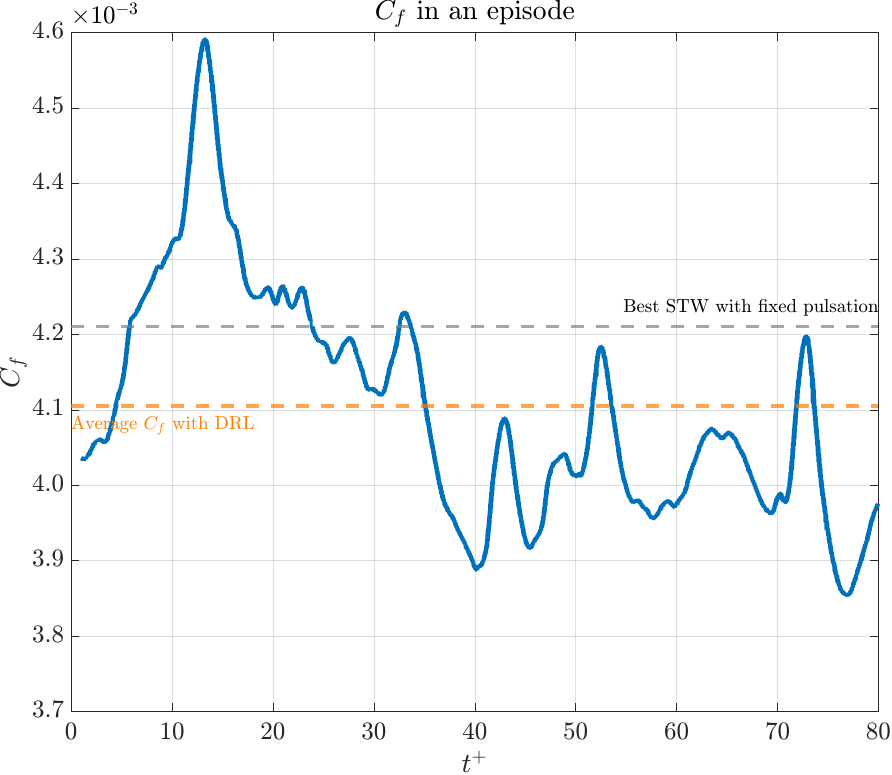}
			\caption{Time evolution of $C_f$ with the trained network, using 8 actuations per episode and implementing the best-performing strategy for $\omega$.}
			\label{fig:cf_ep} 
	\end{minipage}
	\hfill
	\begin{minipage}{.48\textwidth}
		\centering
			\includegraphics[width=1\linewidth]{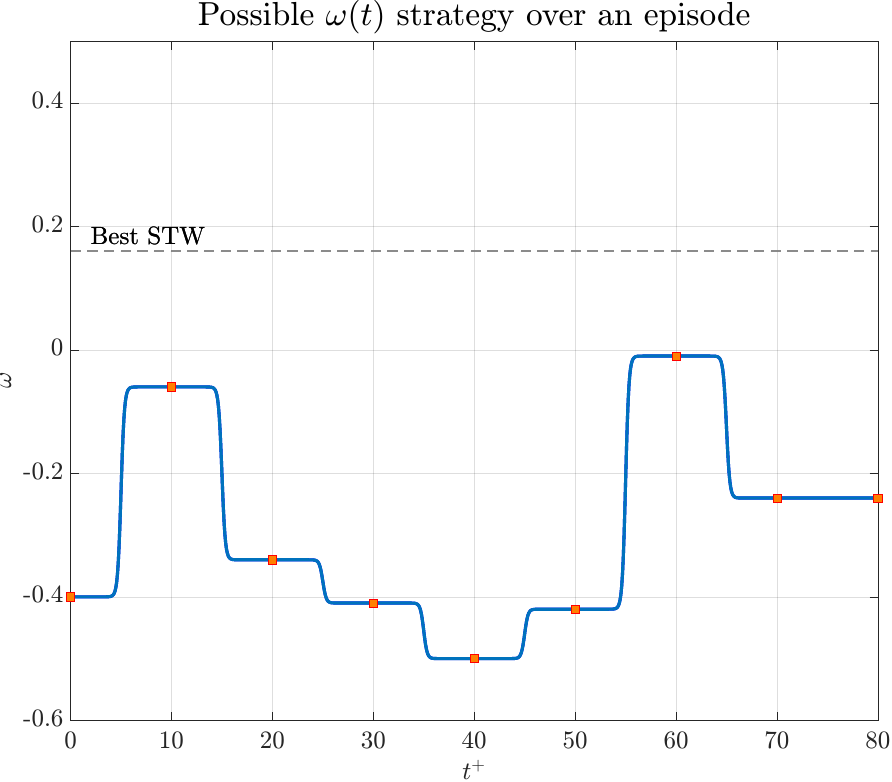}
			\caption{Variation of $\omega$ during one of the episodes corresponding to the best $C_f$ performance shown in Figure \ref{fig:cf_ep}.}
			\label{fig:claw}
	\end{minipage}
\end{figure}

The initial training session, with the reward \(r\) based on the skin friction coefficient \(C_f\) (see \ref{reward1}), focused on achieving an amount of skin friction drag reduction (DR) comparable to the results by Quadrio et al. \citep{quadrioStreamwisetravellingWavesSpanwise2009}. This goal was met, as shown in figure \ref{fig:cf_ep}.
However, the strategies developed were only effective for \(t^+_{\text{ep}} = t_{\text{ep}}U_b/H \approx 80\) time units. We are currently investigating the cause of this limitation, with the long-term objective of controlling the flow for an arbitrary duration.

Figure \ref{fig:claw} shows the time history of the actuation \(\omega(t)\) applied during this first training session. It is noticeable that the values tend to be negative, unlike the best-performing \(\omega\) obtained by Quadrio et al. Furthermore, when compared with their results, it seems that a mild improvement in DR is achieved. However, this conclusion should be taken with caution since the average field may not have reached statistical convergence across the episodes.

The subsequent experiment aimed at maximising the ratio \(\tilde{E}_{ks}/\tilde{E}_k\), focusing on the enhancement of the distribution of the streamwise velocity fluctuations prioritising the presence of straight  near-wall velocity streaks. The latter are characterised as coherent streamwise, alternating positive and negative fluctuations regions that mark the region adjacent to the wall. They typically span about \(100^+\) in the spanwise direction and extend roughly \(1000^+\) in the streamwise length. Their disruption and regeneration are key to the self-sustaining cycle of wall turbulence, as described by \cite{jimenezAutonomousCycleNearwall1999}. During the cycle, periods of low skin friction coefficient coincide with straight streaks undergoing viscous diffusion, while bursts and undulations of the streaks are indicative of periods of high skin friction drag. Therefore, the motivation driving this series of experiments was based on the hypothesis that maintaining the streaks in a straight configuration could inhibit or reduce their instability and, indirectly, reduce skin friction.

\begin{figure}[h!]
	\centering
	\begin{minipage}{.48\textwidth}
		\vbox{
		\centering
			\includegraphics[width=1\linewidth]{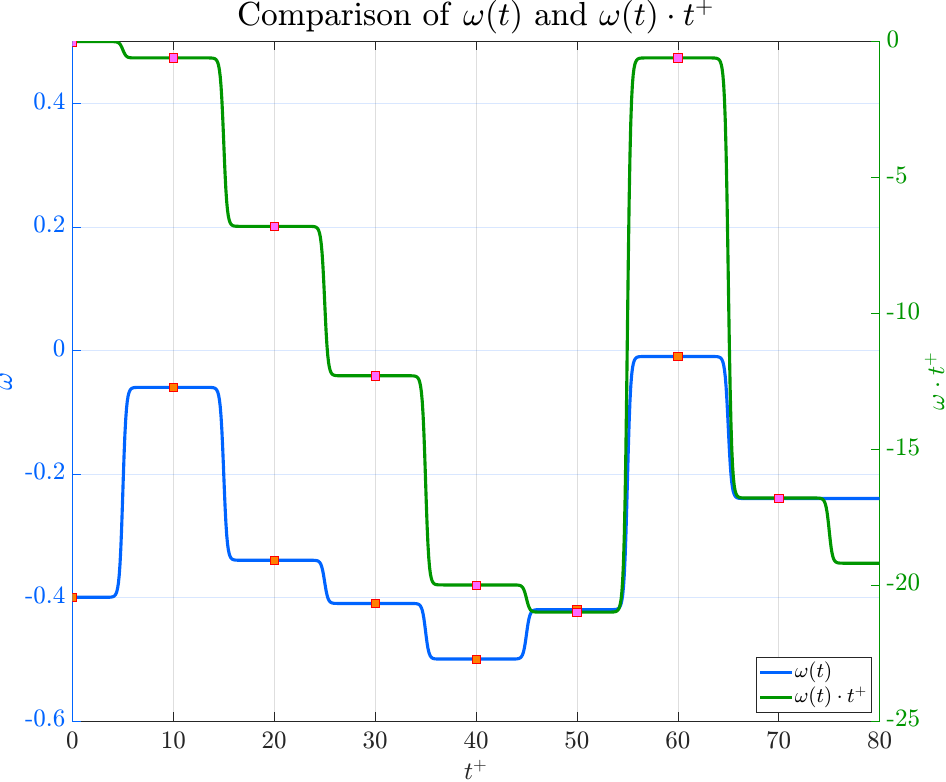}
			\caption{Comparison between the behaviour of $\omega$, in blue, and the value of the phase $\phi = \omega(t)t^+$, in green, with values reported on the right. It is possible to notice that the phase tends to get more and more negative during the episode, until a bump towards the end of it that makes it assume a similar value to the starting one. However, it never exceeds it since the values of $\omega$ remain negative.}
			\label{fig:omega_phi} 
			\vspace{\fill}
        }
	\end{minipage}
	\hfill
	\begin{minipage}{.48\textwidth}
		\vbox{
		\centering
			\includegraphics[width=1\linewidth]{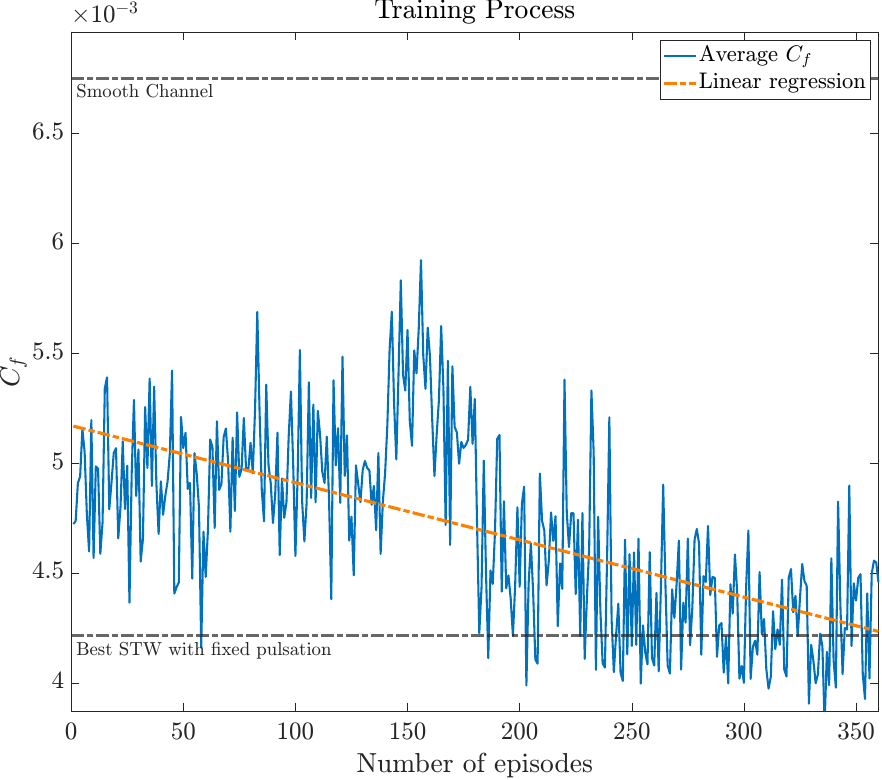}
			\caption{Training process to minimise $C_f$. The horizontal axis corresponds to the number of episodes of the training, while on the vertical axis the $C_f$ value, averaged over each episode, is shown. The network starts learning better strategies after 180 episodes, until it reaches a plateau, after which it is able to outperform the standard STW. The training is overall quite linear, never exceeding the $C_f$ of a case with non-actuated wall condition.}
			\label{fig:training_cf} 
			\vspace{\fill}
        }
	\end{minipage}
\end{figure}

The exploration of the effects of this alternative reward was based on a set of modified parameters applied to both the DNS solver and the DRL libraries. Table \ref{tab:cfd}, right column, summarises those modifications. %These were not limited to the ML libraries but also the the DNS parameters, in
In particular, we have now moved the plane used for data collection to $15^+$, while previously it was $\approx 1^+$. This location should provide richer information as compared to the previous choice
that mainly reflects the state of the laminar sub-layer.
 
Moreover, $y^+=15$ is considered to be an optimal choice for observation planes in DR and flow control feedback \citep{hammondObservedMechanismsTurbulence1998}\citep{chungEffectivenessActiveFlow2011}. 
However, it should also be mentioned that strategies based on sampling at $y^+=1$ remain still appealing due to the possibility of validating with experiments using non intrusive measurements of velocity and stresses at the wall.

\begin{figure}[h]
	\centering
	\begin{minipage}{.48\textwidth}
		\vbox{
		\centering
			\includegraphics[width=1\linewidth]{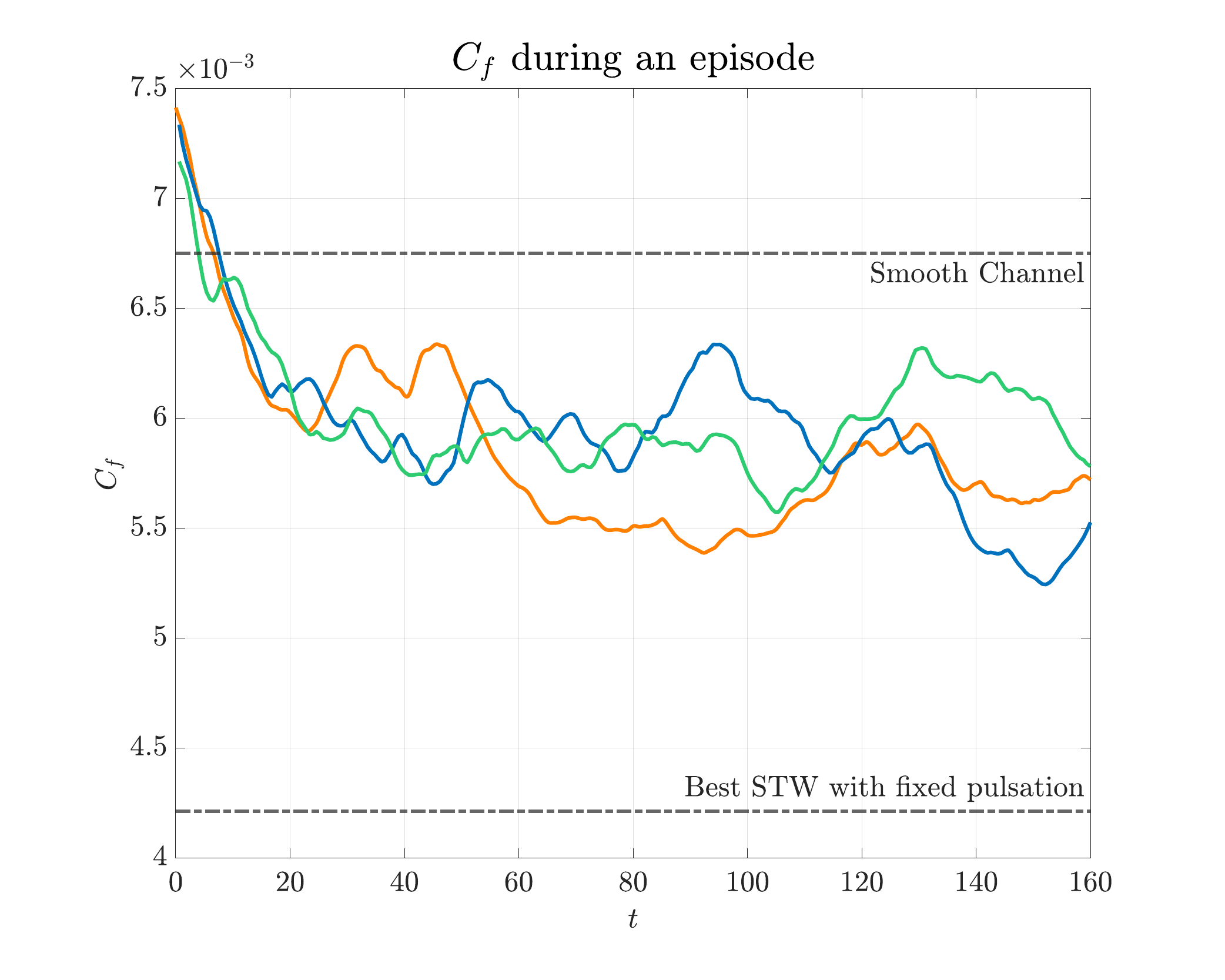}
			\caption{Training of the DRL network - 16 actuations per episode, with maximisation of streamwise velocity fluctuations as a reward (see equation \ref{eq:rew2}). Different colors refer to different runs, which converge to the same value od $\text{d}p/\text{d}x$.}
			\label{fig:cf_ek} 
			\vspace{\fill}
        }
	\end{minipage}
	\hfill
	\begin{minipage}{.48\textwidth}
		\vspace{-20pt}
		\vbox{
		\centering
			\includegraphics[width=1\linewidth]{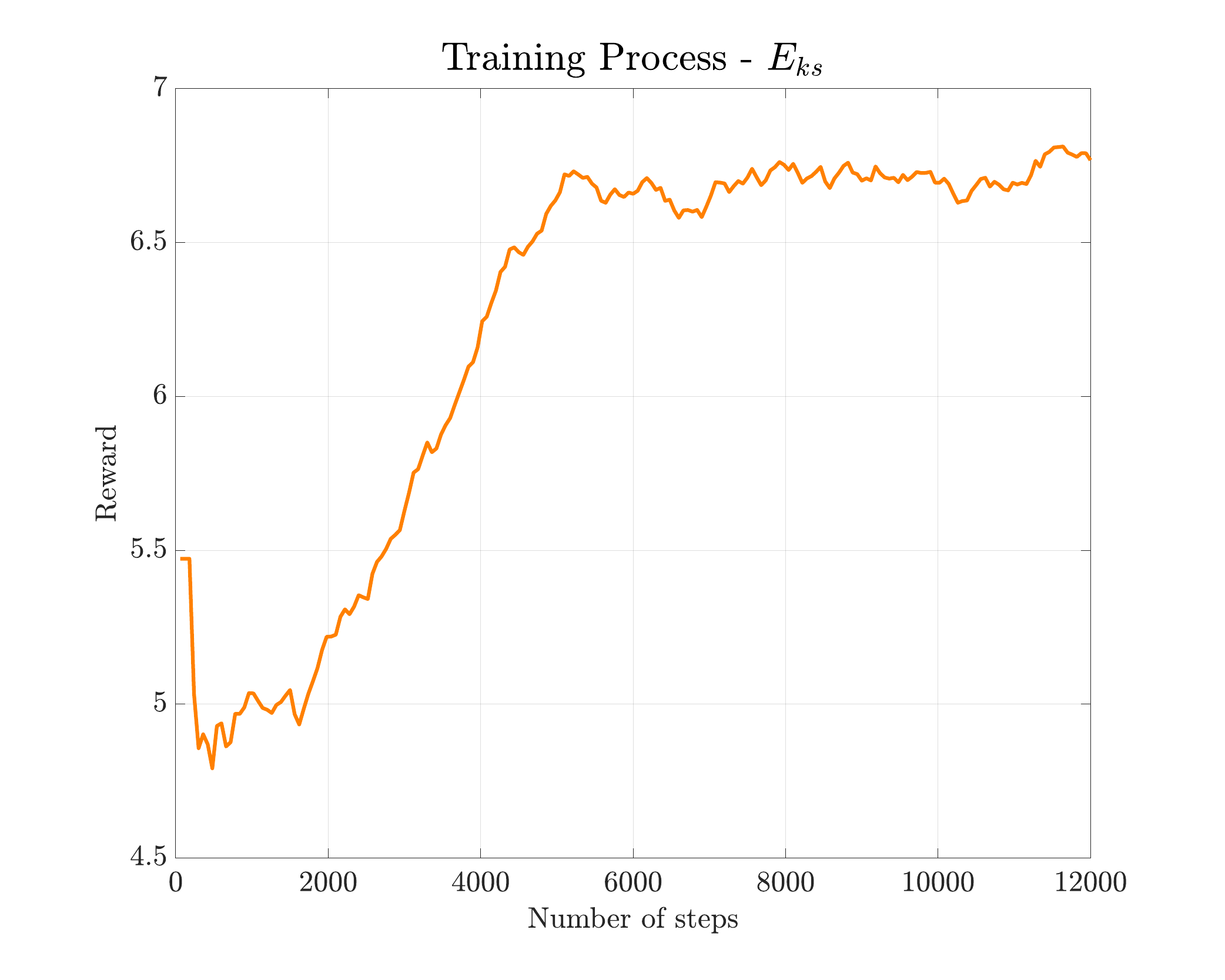}
			\caption{Reward computed as $\tilde{E}_{ks}/\tilde{E}_k$ per training step. The training reaches a plateau at a value higher than the standard STW.}
			\label{fig:training_ek} 
			\vspace{\fill}
        }
	\end{minipage}
\end{figure}

Regarding the results obtained when pursuing the maximisation of the streaks energy content, from the time history of $C_f$ during an episode, reported in figure \ref{fig:cf_ek}, two conclusions can be drawn. Firstly, that the we do obtain a significant \textit{indirect} drag reduction although limited to 
$20\%$  (against the $45\%$ of Quadrio et al.). Secondly, that the episode duration seems to be sufficient to condition the short-term, transient  behaviour of the flow, but not
long enough to establish a consistent drag reducing scenario.

It is also noticed that the value of the reward approaches $r=6.7$, as shown in figure \ref{fig:training_ek}. For this reward value, the DRL policy is not able to reach values of the ratio $\tilde{E}_{ks}/\tilde{E}_k$ greater than $42\%$, with the upper bound not known yet for this kind of control. 
This limitation may indicate the cause-and-effect relationships involved in the regeneration cycle and suggests that other integral quantities might be more effective in manipulating the flow behaviour (e.g., sweep events may not be controlled by the streaks' topology).

\begin{figure}[ht]
	\centering
	\begin{minipage}{.46\textwidth}
		\centering
			\includegraphics[width=1\linewidth]{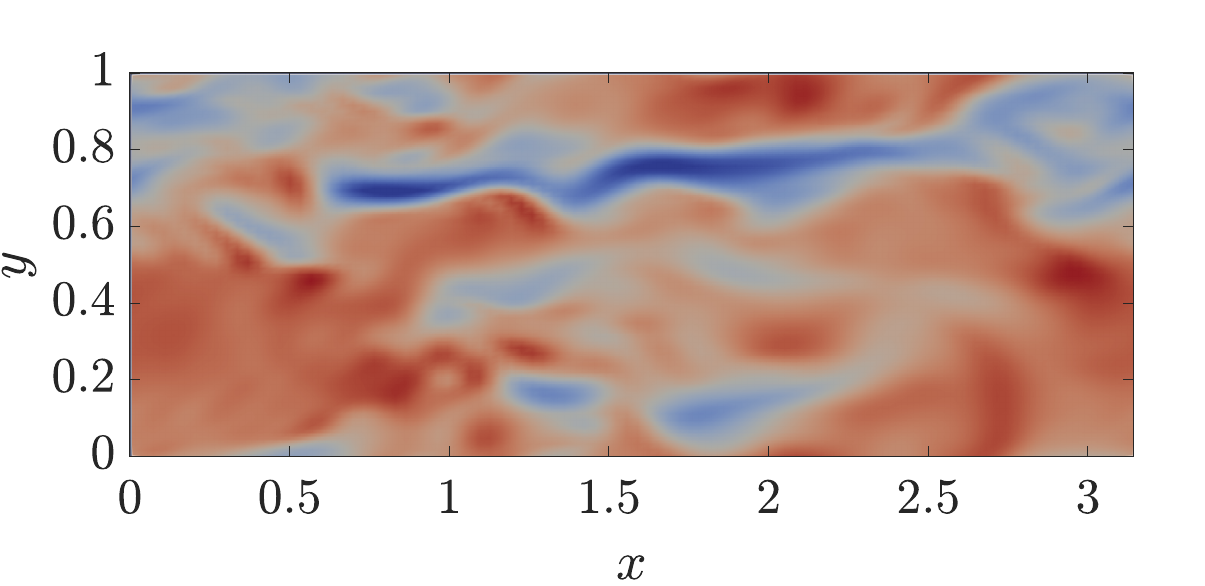}
			\caption{$y^+=20$, non-actuated channel\\ \centerline{ $\tilde{E}_{ks}/\tilde{E}_k=33\%$}}
			\label{fig:not_actuated} 
	\end{minipage}
	\hfill
	\begin{minipage}{.46\textwidth}
		\centering
			\includegraphics[width=1\linewidth]{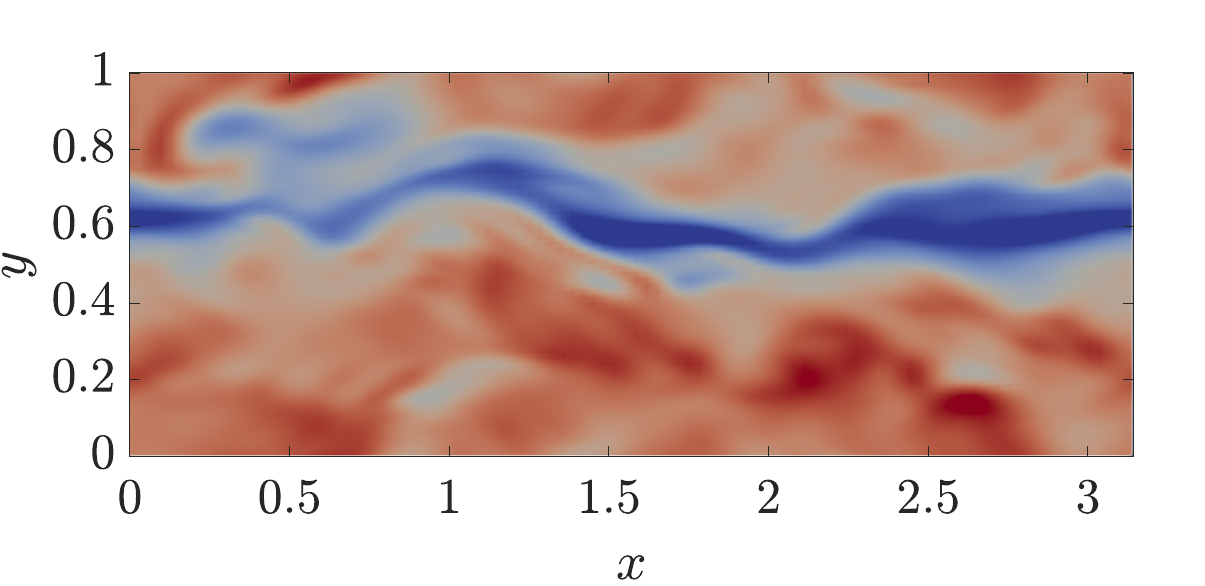}
			\caption{$y^+=20$, $\omega(t)$ STW-DRL control\\ \centerline{ $\tilde{E}_{ks}/\tilde{E}_k=42\%$}}
			\label{fig:actuated} 
	\end{minipage}
	\hfill
	\begin{minipage}{.06\textwidth}
		\centering
			\vspace*{-32pt}
			\includegraphics[width=1\linewidth]{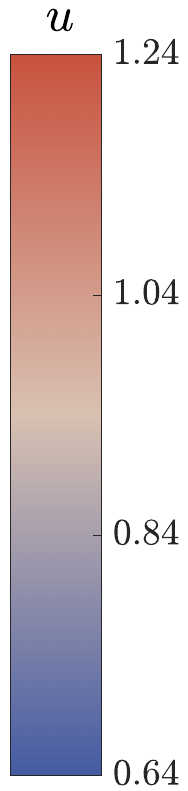}
	\end{minipage}
\end{figure}

Nonetheless, the instantaneous streamwise velocity snapshots extracted on a wall-parallel plane at $y^+=20$ for the actuated and non-actuated cases, shown in figures \ref{fig:not_actuated} and \ref{fig:actuated}, 
show that the maximisation of the energy content of the streamwise velocity fluctuations does have an impact on the flow topology. In particular, the actuated case features a smoother behaviour with less entanglement between low and high speed velocity streaks that present a higher, coherent intensity.

\section{Discussion and conclusion}\label{sec13}
The availability of a code featuring a clear, customisable interface enables the proposal of dynamic flow control techniques. Apart from the technological benefit, the same tool can be used to enhance or dampen the effect of various coherent structures embedded in the wall region, opening new avenues in understanding the wall cycle and its manipulation. Here, we presented a case that focused on the fluctuating energy distribution, but other rewards can be envisaged, such as the minimisation of the fourth quadrant events (i.e., \textit{sweeps} \citep{hammondObservedMechanismsTurbulence1998}) in the streamwise-wall-normal velocity fluctuations distribution.

Although we have presented only two physical realisations, the primary focus of the present contribution is on the implementation of DRL control techniques within the framework of direct numerical simulations. In the context of this work, several key features and choices need to be highlighted and further explored.

\subsection{Actuation duration}

One of the most important parameters to set concerns the time period between consecutive $\omega(t)$ adjustments (see \ref{fig:claw}). 
Initially, the time interval between actuations $\Delta \tau^+$ was set to $10^+$. This choice corresponds to a quarter of the standard STW period when $\omega$ is fixed in time in the maximum DR configuration. It was assumed that a possible control law defined by the DRL algorithm would require an actuation update on a similar time scale. Note that $\Delta \tau^+=10$ implies up to four actuations on  $\omega$ per period of the standard STW, thus giving sufficient relaxation time to the near wall turbulence.

Although we were able to find some interesting results, the choice of a piecewise constant actuation over a certain \(\Delta \tau^+\) may result in a too restrictive control law that does not allow for the determination of the optimal \(\omega(t)\).

The duration of the actuation period can be optimised based on physical considerations. We could start by defining an event \(\mathbb{E}\) as a general change in the topology of the structures within the logarithmic layer. The likelihood of this event occurring depends on the conditions at the wall a few time units prior. Ideally, this delay should be slightly shorter, but not too much shorter, than the \(\Delta \tau^+\) chosen for the actuation period. This would ensure that the action imposed by the STW within the actuation period would have an impact on the close-to-the-wall structures. The agent would also learn faster because the correlation between its actions and the change in the environment would be more direct.

\subsection{Hyper-parameter tuning}
When tackling increasingly complex optimisation problems, it is important to tune the DRL hyper-parameters. \textit{Hyper-parameters} are parameters that can be adjusted to modify the learning behaviour and the agent's features, such as the learning rate and batch size. These hyper-parameters 
can be adjusted depending on the agent's policy without altering the DNS environment. However, their impact on training cannot usually be assessed beforehand, necessitating iterative processes to develop and refine a meaningful policy. This process is time-consuming and difficult to estimate in advance. Future studies will include this tuning to avoid sub-optimal policies and ensure convergence toward a satisfactory reward.

\subsection{New computational architectures}
The numerical simulation of turbulence still predominantly implies the use of CPU-based computer architectures for a series of reasons previously mentioned in the introduction. GPU-based solvers started to appear in the last 
two decades, with outstanding performances when applied to fully explicit algorithms typical of compressible flow solvers where
there is no need for the solution of a pressure Poisson equation. GPU computing for incompressible solvers is more recent
although it has been proven to have the potential to exceed the performances of standard CPU-based solvers \citep{karpLargeScaleDirectNumerical2022}. 

Since DRL libraries based on Python are already routinely used on GPUs architecture and the DNS environment is based on CaNS, that offers compiling instructions to work for GPUs architectures, we plan to migrate all the developed software on GPUs or hybrid GPU-CPUs platforms. This should enhance the performances of the overall numerical methodology which will become of paramount importance
when using a multi-agent approach in the framework of the previously introduced spanwise strips.

\subsection{Conclusion}
This study was motivated by the growing opportunities arising from recent algorithms advancements in deep reinforcement learning (DRL) and deep learning. The DRL methodology and related tools are  promising candidates for improving flow control techniques by enabling actuations to be governed by policies derived from training a DRL agent. They can also be used to explore the underlying physics of wall turbulence by performing numerical experiments that enhance the presence and the interactions between flow structures. This task may be pursued more efficiently with other types of actuations different from the ones we are currently testing.

\section*{Declarations}

\begin{itemize}
\item Funding: R.V. was supported by ERC Grant No. 2021-CoG-101043998, DEEPCONTROL. The views and opinions expressed are however those of the author(s) only and do not necessarily reflect those of European Union or European Research Council. G.C. would like to acknowledge the support of the EPSRC through their DTP studentships programme, grant number EP/W524608/1.
\item Conflict of interest: The authors report no conflict of interest.
\item Data access assessment: The relevant data has been provided in the article, but further data can be provided upon request. The repository to access the DRL-DNS interface embedded in CaNS can be found at the following link: \url{https://github.com/gmcavallazzi/CaNS_DRL}
\end{itemize}

\newpage
\begin{appendices}
\section{Training algorithm - single step}\label{secA1}
\begin{algorithm}[]
\KwIn{$action$} \KwOut{$observation,\,reward,\,flag$}
Check $act$\tcp*{$act$ is the actuation number, starting from 0}
\eIf{$act==0$}
   {send CaNS START instructions\;}
   {send CaNS CONTN instructions\;}
 
Map $action \in [-1, 1] \mapsto [\omega_{min}, \omega_{max}]$ \tcp*{This is needed to rescale $\omega$ within boundaries imposed by the user. $action$ comes from StableBaselines3 routines used when the NN weights are updated.}
Send $\omega$ to CaNS\;
Wait for NS eqs to be solved for $t_{step}$\tcp*{time between actuations}
Receive observations $obs$ from CaNS\;
Receive $rew_{temp}$ from CaNS\tcp*{$rew_{temp}$ is a quantity meaningful to judge the state obtained in the DNS, like $C_f$}
Convert numpy array $obs$ into a grey-scale image\tcp*{Needed with CNNs}
Compute reward $rew = f\left(rew_{temp}\right)$\tcp*{$rew \in [0,1]$}
Update $act=act+1$\;
Update $count=count+1$\tcp*{$count$ keeps track of all the steps performed across all the episodes}
\eIf{$act==n_{act}$}
   {$flag$ = episode finished\; \eIf{$count==n_{tot}$}
   {send CaNS ENDED instructions\tcp*{$n_{act}$ is the number of actuations in an episode, $n_{tot}$ is the total number of timesteps to complete all the episodes}}
   {check, possible I/O issues\;}}
   {$flag$ = episode not finished\; send CaNS CONTN instructions\;}
\caption{Schematised step of DRL training}
\label{alg:step}
\end{algorithm}

\end{appendices}

\bibliography{biblio}% common bib file

\end{document}